\documentclass[numberedappendix]{emulateapj}

\usepackage{xspace, color}
\usepackage[breaklinks,colorlinks,citecolor=blue]{hyperref}

\shorttitle{Accelerated Expansion}
\shortauthors{Rubin and Hayden}

\begin{document}

\title{Is the expansion of the universe accelerating? All signs point to yes}

\author{
D.~Rubin\altaffilmark{1,2} and B.~Hayden\altaffilmark{2,3}
}

\altaffiltext{1}{Space Telescope Science Institute, 3700 San Martin Drive, Baltimore, MD 21218}
\altaffiltext{2}{E.O. Lawrence Berkeley National Lab, 1 Cyclotron Rd., Berkeley, CA, 94720}
\altaffiltext{3}{Department of Physics, University of California Berkeley, Berkeley, CA 94720}

\newcommand{\hubble}{\textit{Hubble Space Telescope}\xspace}
\newcommand{\nsig}{$3.1\sigma$\xspace} 
\newcommand{\nsigdecel}{$4.2\sigma$\xspace} 
\newcommand{\nsigflat}{$8.7\sigma$\xspace} 
\newcommand{\nsigdecelflat}{$11.2\sigma$\xspace}

\email{drubin@stsci.edu}

\begin{abstract}
The accelerating expansion of the universe is one of the most profound discoveries in modern cosmology, pointing to a universe in which 70\% of the mass-energy density has an unknown form spread uniformly across the universe. This result has been well established using a combination of cosmological probes \citep[e.g.,][]{planck}, resulting in a ``standard model'' of modern cosmology that is a combination of a cosmological constant with cold dark matter and baryons. The first compelling evidence for the acceleration came in the late 1990's, when two independent teams studying type Ia supernovae discovered that distant SNe Ia were dimmer than expected. The combined analysis of modern cosmology experiments, including SNe Ia, the Hubble constant, baryon acoustic oscillations, and the cosmic microwave background 
has now measured the contributions of matter and the cosmological constant to the energy density of the universe to better than 0.01, providing a secure measurement of acceleration. A recent study \citep{nielsen16} has claimed that the evidence for acceleration from SNe Ia is ``marginal.'' Here we demonstrate errors in that analysis which reduce the acceleration significance from SNe Ia, and further demonstrate that conservative constraints on the curvature or matter density of the universe increase the significance even more. Analyzing the Joint Light-curve Analysis 
supernova sample, we find \nsigdecel evidence for acceleration with SNe Ia alone, and \nsigdecelflat in a flat universe. With our improved supernova analysis and by not rejecting all other cosmological constraints, we find that acceleration is quite secure.

\end{abstract}

\section*{Introduction}

The discovery of the accelerating universe by two teams \citep{R98,P99} in the late 1990's was one of the major breakthroughs in cosmology. Using type-Ia supernovae (SNe~Ia) as standard candles, both teams independently determined that high-redshift SNe were fainter than expected in a matter-dominated universe, implying the need for a cosmological constant (or more generally, dark energy) to accelerate the expansion of the universe, increasing cosmological distance as a function of redshift. 

SNe Ia are not perfect standard candles, however. Work leading up to the discovery \citep{Phillips93,Riess96,hamuy96,Perlmutter97} demonstrated the need for empirical standardization relations. Peak absolute magnitudes correlate with the width of the light curve (broader-light-curve SNe are more luminous) and the color of the supernova (redder SNe are less luminous). In the years since, other empirical standardization relations have been noted, including one related to host-galaxy stellar mass \citep{kelly10,sullivan10} (perhaps driven significantly by the local star formation rate, \citealt{rigault13}).

The Sloan Digital Sky Survey (SDSS) and SuperNova Legacy Survey (SNLS) SN teams have completed the Joint Light-curve Analysis (JLA) \citep{JLA}. This analysis incorporates a thorough recalibration of both surveys \citep{betoule13}, and the full set of spectroscopically confirmed SDSS SNe Ia \citep{sako14}; it represents the most up-to-date large SNe Ia compilation.\footnote{Currently, JLA represents $\sim 60\%$ of the world sample of SNe Ia, so we expect constraints to continue to rapidly improve.} A recent claim \citep[][hereafter N16]{nielsen16} was made that this dataset provides only ``marginal evidence'' for acceleration. We examine the statistical model N16 used to make this claim, and find it deficient for the task. In particular, a simple (and well-justified) update of the model to better account for changes in the observed SN light-curve parameter distributions with redshift significantly increases the statistical strength of the acceleration evidence.

\section*{The Statistical Model}

In the case of JLA, the standardization relations employed are light-curve width ($x_1$ in the framework of SALT2,  \citealt{guy07}), color ($c$), and host-galaxy stellar mass. The dependent variable is taken to be the rest-frame $B$-band magnitude ($m_B$). The light-curve parameters are determined by comparing a rest-frame spectral energy distribution model to the observer-frame photometry; similarly, the host stellar mass is estimated from broad-band photometry. The cosmological results rely on the ability of the statistical framework to fit the standardization relations (in JLA, these are taken to be linear in $x_1$ and $c$, and a step function in host mass), yet the uncertainties (a general term that we take to include unexplained dispersion around the model) in the dependent variable ($m_B$) and independent variables ($x_1$, $c$, host mass) are of similar size. The JLA analysis itself used a frequentist line-fitting procedure with only modest biases in its regime of applicability \citep{mosher14}. 

In contrast, the statistical model from N16 uses a Bayesian Hierarchical Model  (c.f., \citealt{gull89}). In the N16 model, the latent (``true'') parameters for each SN are modeled with nuisance parameters, which are marginalized over to obtain inference on the global parameters. The distribution of the latent parameters must be adequately described by the prior. For example, flat priors on the latent variables cause a bias in the fit \citep{gull89}. Making the parameters of the prior (``hyperparameters'') part of the model avoids this bias (this multi-level nature gives rise to the name ``Hierarchical'').

The key shortcoming of the N16 model is that it assumes redshift-independent distributions for $x_1$ and $c$. As shown in Figure~\ref{fig:JLAvsz}, the observed distributions (plotted points) are far from redshift-independent. Two effects visible in the data---selection effects and the correlation between older host galaxies and narrower-light-curve (lower $x_1$) SNe \citep{hamuy95}---result in a more luminous distribution of SNe as an increasing function of redshift. The selection effects are particularly evident in color, where only bluer SNe (more negative $c$) are above the completeness limit for the high-redshift end of each ground-based sample. By incorrectly treating these distributions as redshift-independent, N16 biased their latent $x_1$ and $c$ towards the global mean, effectively removing some of the standardization \citep{conley07, woodvasey07, karpenka15}.\footnote{N16 claim their model is not Bayesian, and use frequentist inference for some global parameters, but the marginalization in their Equation 8 is a Bayesian approach. The priors assumed thus affect the inference.} The JLA sample was corrected for selection bias in \citet{JLA}, but only in the sense that SNe which are selected to be more luminous {\it after standardization} are adjusted to be less luminous. The bias correction cannot compensate for a deficient standardization, as provided by a constant-in-redshift model of the distributions.

\subsection*{Redshift-Independent Distributions}

As a starting point, we perform an analysis similar to that in N16, using Hamiltonian Monte Carlo to sample from the posterior (we describe the details in Appendix~\ref{sec:app}). We assume a cosmological model with cold matter and a cosmological constant ($\Lambda$CDM). We make four measurements: a $\Lambda$CDM universe allowed to have spatial curvature (i.e., $\Omega_m + \Omega_{\Lambda} \neq 1$), and a flat $\Lambda$CDM universe (the assumption of flatness is discussed more in the discussion section), each with both sets of model assumptions. We compute the deceleration parameter $q_0$, ($q_0 \equiv \left. -\frac{\ddot{a}}{a\,H^2} \right|_{t = t_0}$, equal to $\Omega_m/2 - \Omega_{\Lambda}$ for a $\Lambda$CDM cosmology). We evaluate the statistical significance of acceleration ($q_0 < 0$) by comparing the 50th percentile of the posterior with the difference of the 50th percentile and the 84th percentile (taken as an estimate of 1$\sigma$ in the $+q_0$ direction), and then rounding to $0.1\sigma$. The statistical significance of the acceleration is \nsig with no constraint on curvature, and \nsigflat assuming a flat universe (see Figures \ref{fig:OmOL}, \ref{fig:qnot}, left panels). The difference between this estimate and one derived from explicitly measuring the fraction of samples with $q_0 > 0$ is modest (66 posterior samples out of 60,000 have $q_0 > 0$).

\subsection*{Redshift-Dependent Distributions}
\begin{figure*}[ht]
\centering
\includegraphics[width=\linewidth]{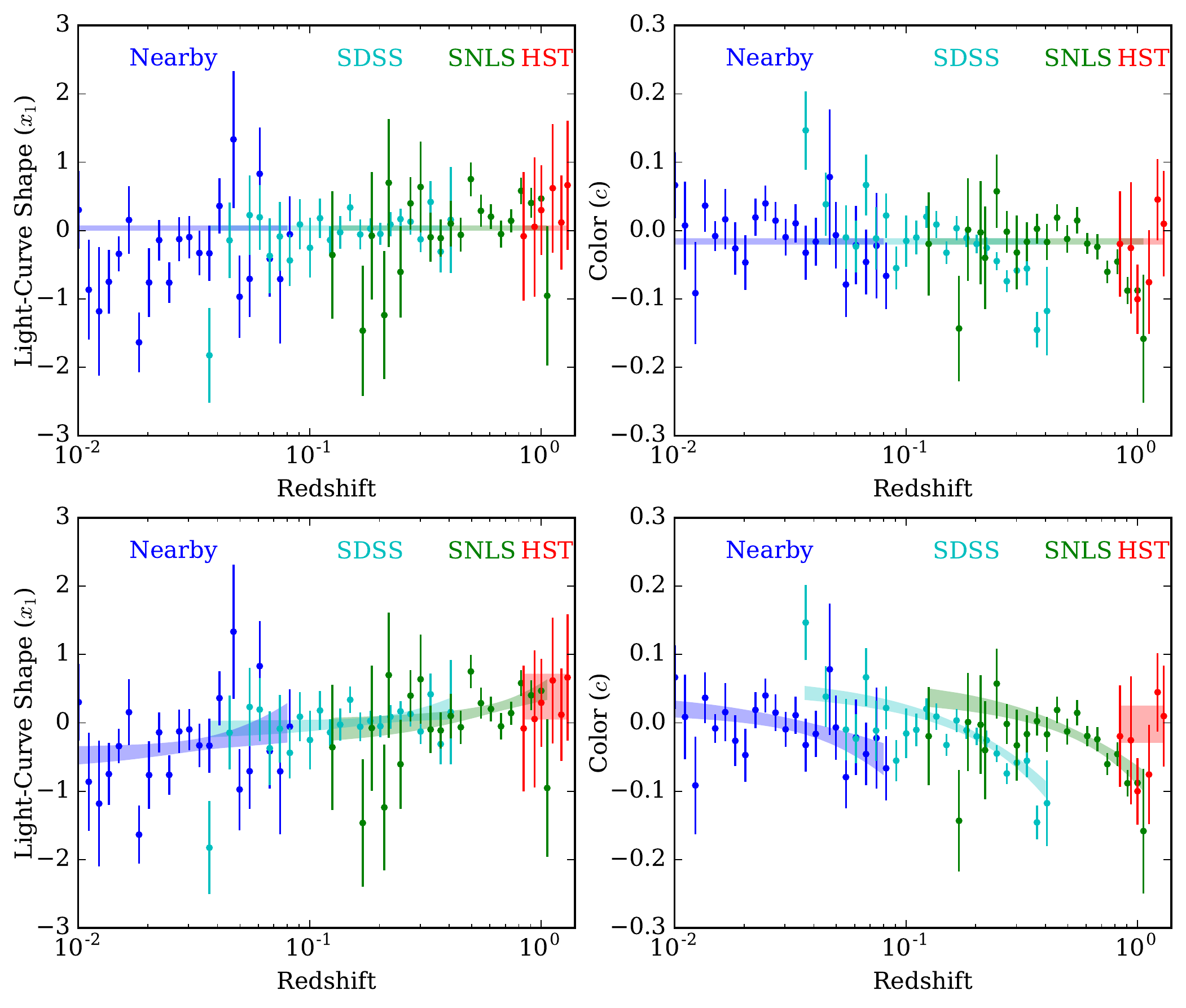}
\caption{Binned $x_1$ (left panels) and $c$ (right panels) light-curve parameters as a function of redshift for the JLA sample. The trend of color with redshift within each ground-based sample is expected due to the combination of the color-luminosity relation combined with redshift-dependent luminosity detection limits. The top panels show the 68\% credible constraints on a constant-in-redshift model, as was used in N16. The bottom panels show our proposed revision. Failing to model the drift in the mean observed distributions demonstrated by the bottom panels will tend to cause high-redshift SNe to appear brighter on average, therefore reducing the significance of accelerating expansion.}
\label{fig:JLAvsz}
\end{figure*}

\begin{figure*}[ht]
\centering
\includegraphics[width=\linewidth]{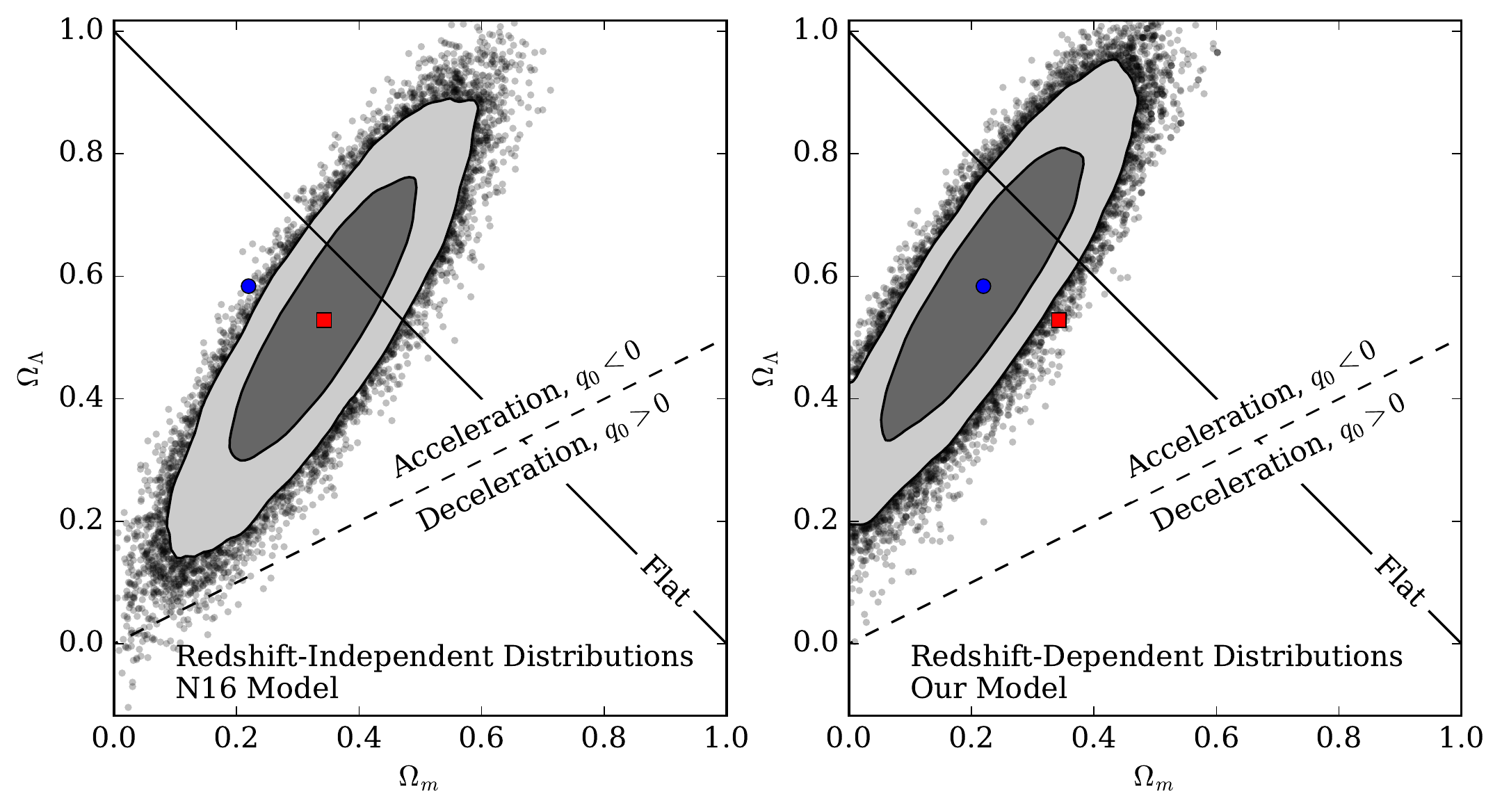}
\caption{$\Omega_m$-$\Omega_{\Lambda}$ constraints enclosing 68.3\% and 95.4\% of the samples from the posterior. Underneath, we plot all samples. The left panel shows the constraints obtained with $x_1$ and $c$ distributions that are constant in redshift, as in the N16 analysis; the right panel shows the constraints from our model. The red square and blue circle show the location of the median of the samples from the respective posteriors.}
\label{fig:OmOL}
\end{figure*}

Next, we introduce a simple model of the observed distributions as a function of redshift. We allow each source of SN discovery (Nearby, Sloan Digital Sky Survey, SuperNova Legacy Survey, \hubble) to have a linear variation in the mean with redshift (for the \hubble SNe, we use only a constant mean in redshift, as this sample is too small to constrain any variation). This model is shown in Figure~\ref{fig:JLAvsz}; the variation with redshift is highly statistically significant. We also try a more flexible model in redshift \citep{rubin15}, and it makes only a small difference \citep[the only requirement on the model is to be at least as flexible in redshift as the cosmological model under consideration,][]{rubin15}. The statistical significance of the acceleration increases to \nsigdecel, and \nsigdecelflat assuming a flat universe (see Figures \ref{fig:OmOL}, \ref{fig:qnot}, right panels). Again, the difference between this estimate and one derived from explicitly measuring the fraction of samples with $q_0 > 0$ is modest (only one posterior sample out of 60,000 has $q_0 > 0$).

\subsection*{Other Cosmological Models}
\begin{figure*}[ht]
\centering
\includegraphics[width=0.75 \linewidth]{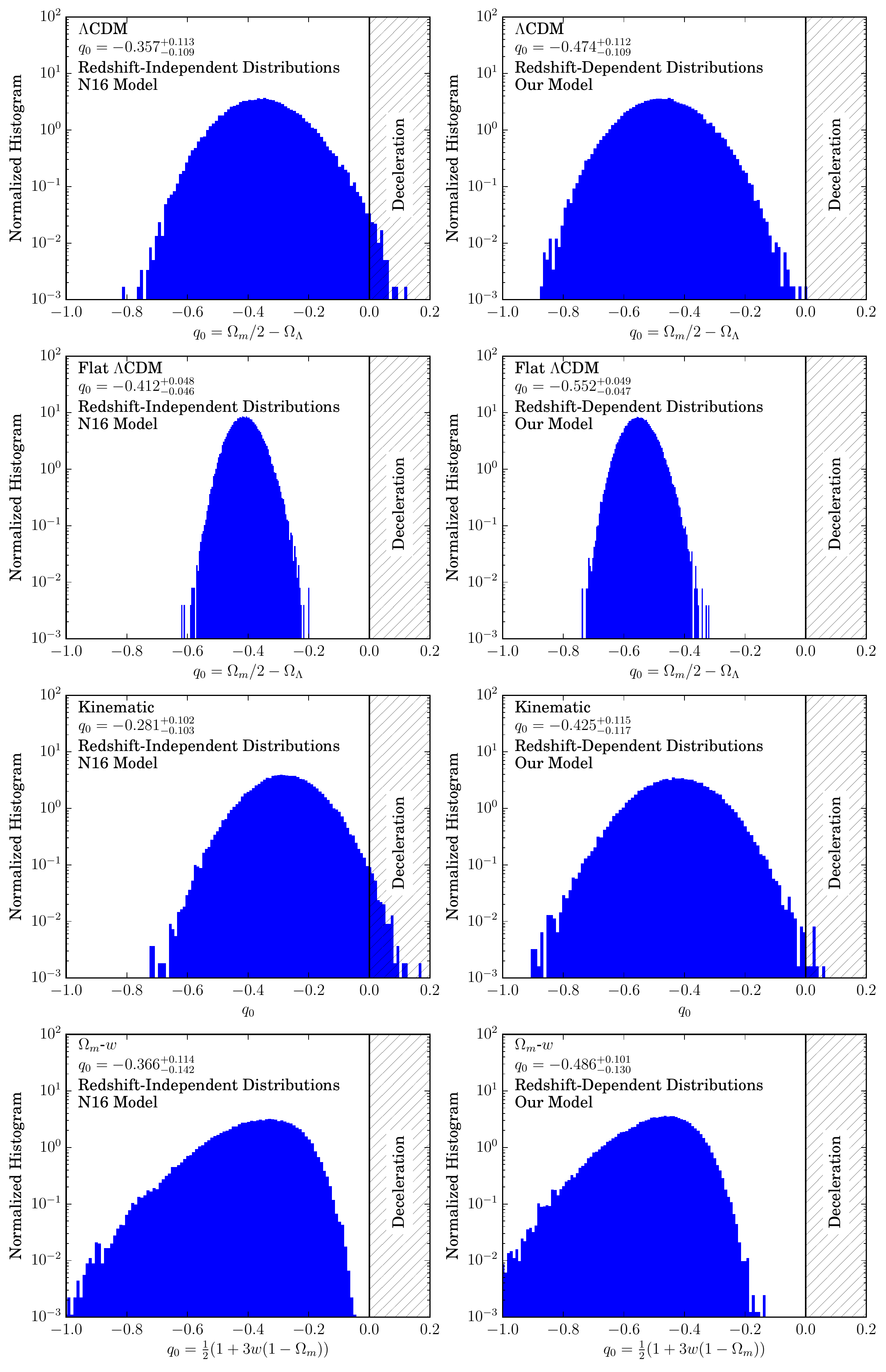}
\caption{$q_0$ histograms, normalized to have an integral of unity. The left panels show the constraints for each cosmology with a constant-in-redshift model of the light-curve-parameter distributions, as in N16; the right panels show our model. In every case, the statistical significance of the acceleration is higher with our redshift-dependent distribution model. The top row shows the results for $\Lambda$CDM cosmologies with curvature allowed. The next row down shows $\Lambda$CDM cosmologies with a flat universe. The next row shows the constraints with the kinematic expansion in redshift. Finally, the bottom row shows the results for $\Omega_m$-$w$.}
\label{fig:qnot}
\end{figure*}

For a result that relies only on kinematics, we also compute $q_0$ constraints using the \citet{visser04} series expansion of luminosity distance as a function of redshift. We take the first three terms (including $q_0$ and $j_0 \equiv \left. \frac{\dddot{a}}{a\,H^3} \right|_{t = t_0}$). The $q_0$ constraints are illustrated on the third row of Figure~\ref{fig:qnot}. Even with flat priors on $q_0$ and $j_0$ (allowing the kinematics to venture into regions of parameter space that may be hard to realize dynamically), we find 3.7$\sigma$ evidence for acceleration (2.8$\sigma$ with the N16 model). SN data alone can be used to derive constraints on the joint posterior of $\Omega_m$ and the dark energy equation of state parameter ($w=P_{\mathrm{DE}}/\rho_{\mathrm{DE}}$), assuming a flat universe \citep{garnavicheqofst, P99}; we compute constraints for this model as well. For simplicity, we take a flat prior on both $\Omega_m$ and $w$, and find strong evidence for $q_0<0$, as shown in the bottom row of Figure~\ref{fig:qnot}. The constraints on $q_0$ are non-Gaussian, but in both the N16 model and ours, no samples (out of 60,000) have $q_0>0$.

\begin{figure}[ht]
\centering
\includegraphics[width=\linewidth]{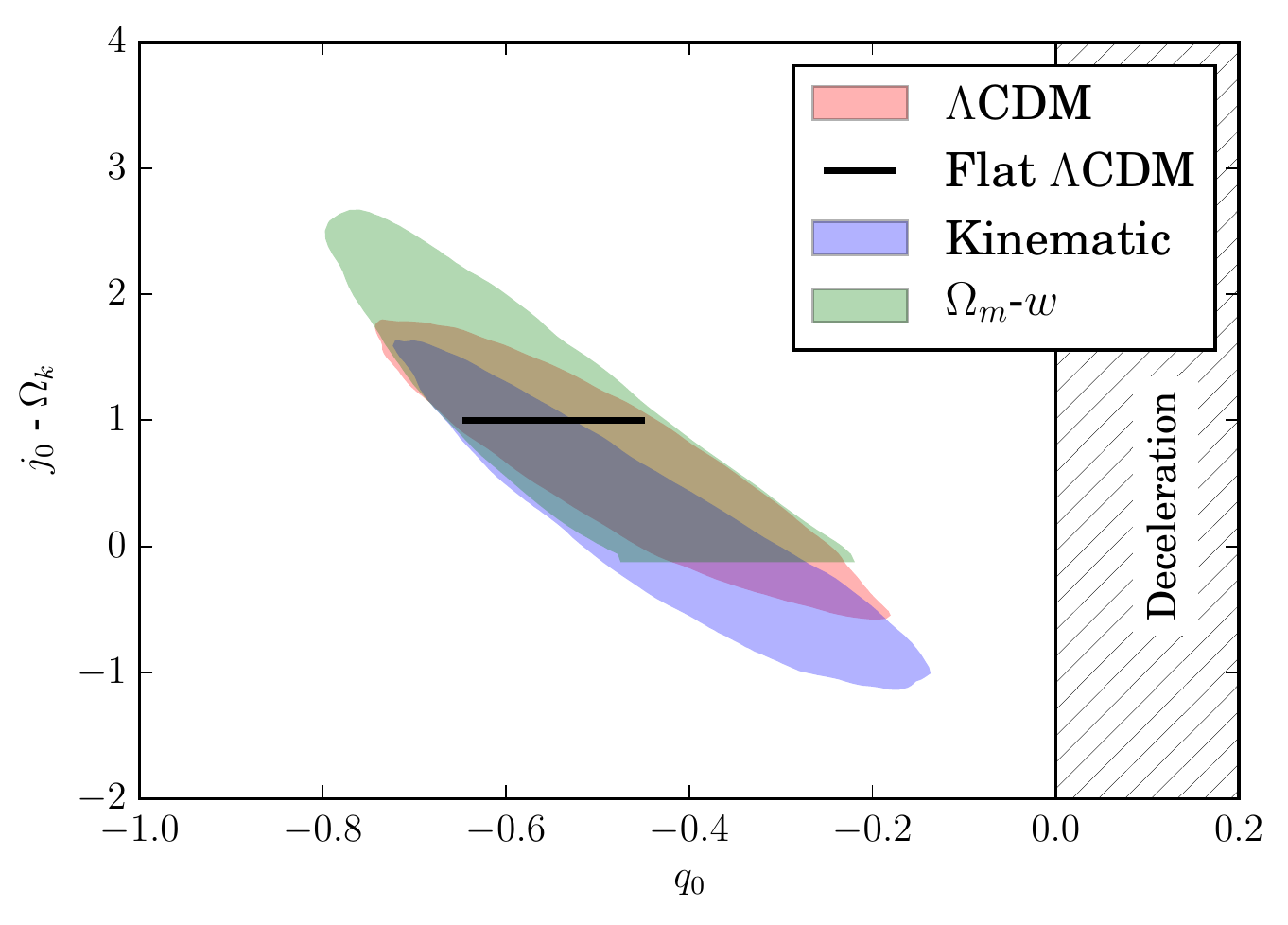}
\caption{$q_0$-$[j_0 - \Omega_k]$ constraints enclosing 95.4\% of the samples from the posterior. For the flat $\Lambda$CDM results, we show the 1D credible interval.}
\label{fig:qnotjnot}
\end{figure}

We next project the constraints from each of our four models to the $q_0$-$[j_0 - \Omega_k]$ plane, shown in Figure~\ref{fig:qnotjnot} (both $j_0$ and $\Omega_k$ contribute linearly at the same order in luminosity distance, so we cannot distinguish them in this plane). The constraints for 2D models (models other than flat $\Lambda$CDM) are similar. However, the high $q_0$/low $[j_0 - \Omega_k]$ region is disfavored by the dynamical models, as $\Omega_m$-$w$ cannot reach $j_0 < -1/8$, and even an empty universe \citep[``Milne Model''][]{milne35} only has $j_0 - \Omega_k = -1$.

\section*{Discussion}

Our results (flat-universe $\Omega_m = 0.298^{+0.033}_{-0.031}$) are similar to the frequentist JLA analysis (flat-universe $\Omega_m = 0.295 \pm 0.034$). This is unsurprising; frequentist and Bayesian analyses will converge to exactly the same results under a set of assumptions not far from those made here \citep{rubin15}. We also note that more advanced analyses can better take into account statistical properties of the data (modeling selection effects, non-linear standardization relations, a redshift-dependent host-mass relation, outliers, and a model of unexplained dispersion incorporating $x_1$ and $c$) \citep{rubin15}. N16 did not include the host-mass standardization; excluding this only has a small impact on our results.\footnote{For $\Lambda$CDM with curvature, the significance of the acceleration changes from $4.2\sigma$ with to $4.3\sigma$ without the host-mass relation.} However, we focus our attention on the N16 model of the $x_1$ and $c$ distributions, as it is this model that drives the difference from the JLA analysis.

While constraints derived from SNe Ia alone require a $\sim 30\%$ flatness constraint to push the supernova measurement of acceleration above $5\sigma$, current experiments have constrained curvature to much better precision than $1\%$ \citep{planck}. Even constraints on $\Omega_m$ \citep[e.g., galaxy clusters,][]{allen11}, which imply $\Omega_m>0.2$, cut off the tail of the SN-only posterior extending down to a Milne universe and $q_0 > 0$, allowing the acceleration to again reach $5\sigma$ confidence.\footnote{N16 generalize the Milne model to mean any universe with $q=0$ at all times, rather than specifically an empty universe. In the $q_0$-$[j_0 - \Omega_k]$ plane (Figure~\ref{fig:qnotjnot}), it is clear why the distinction is important; models that are curvature-dominated allow low $j_0 - \Omega_k$, and thus high $q_0$; even weak flatness constraints disfavor that portion of parameter space.} With the combination of current experiments (SNe Ia, baryon acoustic oscillations, cosmic microwave background, and the Hubble constant), $\Omega_\Lambda$ is constrained to be $0.6911 \pm 0.0062$ \citep{planck}. In order to claim that the evidence for acceleration is ``marginal,'' it is necessary to fully reject all measurements of the curvature of the universe, the basic constraints on the matter density of the universe, and other cosmological datasets.

Even without external constraints, this work demonstrates that a more accurate model for the supernova analysis greatly increases the significance of acceleration. We conclude that the analysis in N16 is both incorrect in its method and unreasonable in its assumptions, leading the authors to question a result that is quite secure when addressed properly.

\section*{Acknowledgements}

We appreciate the feedback we received from Greg Aldering, Peter Nugent, Saurabh Jha, Saul Perlmutter, Alex Kim, Peter Garnavich, and Mike Hobson. Support was provided by the Director, Office of Science, Office of High Energy Physics, of the U.S. Department of Energy under contract No. DE-AC02-05CH11231 and NASA ROSES-14 WFIRST Preparatory Science program 14-WPS14-0050.

\clearpage
\appendix
\section{Sampling from the Posterior} \label{sec:app}

We sample from the posterior using Stan \citep{carpenter16} through PyStan (\url{https://pystan.readthedocs.io}). Following \citet{nielsen16}, we assume flat priors on all parameters, but require $\Omega_m > 0$. Our chains are 2500 samples each (after warmup), and show excellent convergence (the diagnostic of \citealt{gelman92} is smaller than 1.01). We run twenty four chains for the results with curvature, and eight chains for the flat-universe results.

As in \citet{rubin15}, in order to speed up sampling, we decompose the light-curve fit covariance matrix into its eigenvectors and sample over the projections onto these (this results in a covariance matrix that has no correlations between SNe). In addition to a fully Bayesian analysis, the only other change we make from N16 is to include the host-mass standardization, as was done in JLA. We make our code available \citep{code16}.

\end{document}